\documentstyle[12pt]{article}
\begin{document}
\baselineskip 24pt
\begin{flushright}
SAGA-HE-123\\
May  1997
\end{flushright}
\vspace{1.0cm}
\begin{center}
{\Large\bf Do really the seagull terms survive\\
{\hfil \hfil}\\
for the electric polarizability of the  nucleon  ?}
\end{center}
\vspace{0.5cm}
\begin{center}
{\large S. SAITO}\footnote{E-mail address 
saito@nuc-th.phys.nagoya-u.ac.jp}\\
{\sl Department of Physics, Nagoya University, Nagoya 464-01}\\
{\large M. UEHARA}\footnote{E-mail address ueharam@cc. saga-u.ac.jp} \\
{\sl Department of Physics, Saga University, Saga 840}
\end{center}
\vspace{0.5cm}
\noindent{\bf Abstract}\par
The seagull terms for the electric polarizability of the nucleon are 
shown to vanish indeed, if one introduces fluctuations around the 
Skyrmion configuration,  and after all the origin of the electric 
polarizability cannot be  attributed to the seagull terms in the Skyrme 
model. 
\newpage
\newcommand\beq{\begin{equation}}
\newcommand\eeq{\end{equation}}
\newcommand\beqa{\begin{eqnarray}}
\newcommand\eeqa{\end{eqnarray}}
\newcommand\noeq{\nonumber}
\def\disp{\displaystyle}
\def\noi{\noindent}
\def\Ref#1{\,[{#1}]}
\def\Nc#1{O($N_c^{#1}$)}
\def\eps{\varepsilon}
\def\finv{\frac{1}{\fpi}}
\def\fpi{f_\pi}
\def\mpi{m_\pi}
\def\D{\Delta}
\def\Lam{\Lambda_S}
\def\half{\frac{1}{2}}
\def\pha{\Phi_a}
\def\phb{\Phi_b}
\def\phc{\Phi_c}
\def\phd{\Phi_d}
\def\ph0{\Phi_0}
\def\pia{\pi_a}
\def\pib{\pi_b}
\def\vx{{\bf x}}
\def\der{\partial}
\def\hatr{\hat r}
\section{Introduction}
Electric and magnetic polarizabilties are important structure 
parameters of baryons, which characterize dynamical responses for 
the external electromagnetic fields. There have been many calculations 
on the electric polarizability of the nucleon and the hyperons by chiral 
soliton models\Ref{\ref{1}-\ref{6}}. In these calculations its origin  is 
attributed to the so-called seagull terms proportional to  $(eA_0)^2$ 
appearing in the Hamiltonian.   
\par
It has been pointed out that the existence of the electric seagull terms 
in the Compton scattering amplitude conflict with the gauge 
invariance of the electromagnetic fields\Ref{\ref{Lvov}}. And 
we  have explicitly proved for the Compton scattering amplitude 
in Ref.\Ref{\ref{SUplb} } that  the  seagull terms contributing to  the 
electric polarizability of the nucleon vanish for a wide range of theories 
including the Skyrme model.   Subsequently we  have successfully 
calculated the electromagnetic polarizabilities of \Nc{} from the 
Compton scatteringamplitude through the dispersion integrals of 
the pion photo-production amplitudes in the Skyrme 
model\Ref{\ref{SUprd}}. 
\par
However, Scoccola and Cohen\Ref{\ref{SC}} gave an argument that the 
seagull terms could survive for the electric polarizability in the Skyrme 
model,  if the fields are restricted to the Skyrmion configurations with 
the collective coodinates in the Lagrangian.Another argument by Meier 
and Walliser followed\Ref{\ref{MW}}:   They insist that the electric 
seagull terms  could survive in the Skyrme model, even if the 
fluctuations around the Skyrmion are introduced in the Lagrangian.
 \par
In this paper we  show again that the  seagull terms indeed vanish in the 
Hamiltonian,  if the fluctuation fields around the Skyrmion are 
introduced.   Our conclusions are that the surviving seagull terms of the
collective Hamiltonian are not stable against any fluctuations, and that 
the electric polarizability of the nucleon cannot be attributed to such 
seagull terms in the Skyrme model.
\par
In the next section we difine variables and notations and give the 
collective seagull terms surviving  in the subspace of the 
collective variables. It is shown in the section 3 that if we introduce 
fluctuations around the rotating Skyrmion,  the seagull terms cannot 
really survive. The conclusions and discussion are given in the 
last section. \par
 
\section{Collective seagull terms}
Let us start with the Lagrangian given in \Ref{\ref {SUprd}} and we 
rewrite it in the form suitable to  our purpose, that is the electric 
polarizability in the static external  elctric field  expressed in terms of 
$A_0$;
\beqa
L&=&\int d^3x\left\{\half\dot\pha K_{ab}\dot\phb+eJ_aK_{ab}\dot\phb-
{\cal{V}}[\Phi]+\half(eA_0B_0 + e^2J_aK_{ab}J_b)\right\}\noeq\\
&=&\int d^3x\left\{\half(\dot\pha+eJ_a)K_{ab}(\dot\phb+eJ_b)-
{\cal{V}}[\Phi]+\half eA_0B_0\right\}
\eeqa
where the SU(2) field $U(x)$ is written as\Ref{\ref{HSUnew}}
\beq
U(x)=\finv[\ph0(x)+i\tau_a\pha(x)]
\eeq
under the constraint 
\beq
\ph0^2(x)=\fpi^2-\sum\pha^2(x),
\eeq 
and 
\beq
J_a=A_0\eps_{3ca}\phc,
\eeq
and $B_0$ is the baryon number density, and see
Ref.\Ref{\ref{SUprd}} for the explicit expressions for $K_{ab}$, 
${\cal{V}}[\Phi]$ and others, which are  not used here. Note that there is 
the seagull term $e^2J_aK_{ab}J_b$ in the Lagrangian. 
\par
At first, we reproduce briefly the argument by \Ref{\ref{SC}}, 
where $\pha$'s  are replaced by the Skyrmion configurations with 
rotating collective coordinates in the {\it laboratory system};
\beq
\pha=\psi_a(\vx,t)=R_{ai}(t)\fpi \hatr_i\sin F(r)
\eeq
with $F(r)$ being the chiral angle, $r=|\vx|$ and $\hatr_i=x_i/r$.  
$R_{ai}(t)$ is the time dependent rotation matrix. The time derivative of 
the field is written as 
\beq
\dot\pha(x)=\dot\psi_a(\vx,t)=Z^c_a(\vx,t)\omega_c,
\eeq
where $\omega_c$ is the angular velocity around the isospin  axis $c$, 
and  $Z_a^c$ is the rotational zero-mode function in the laboratory 
system,
\beq
Z_a^c(\vx,t)=\eps_{abc}\psi_b(\vx,t).
\eeq 
The Lagrangian is then written as 
\beq
L_{\rm Sky}=\int d^3x\left\{\half(Z_a^c\omega_c+eJ_a)K_{ab}
(Z_b^d\omega_d+eJ_b)
-{\cal{V}}+\half eA_0B_0\right\},
\eeq
where the arguments in $J_a$, $Z_a^c$, $K_{ab}$ and others are to be 
replaced by $\psi_a$'s. 
\par
We define the intirnsic spin operator conjugate to $\omega_c$ 
as
\beqa
I_c&=&\frac{\der L}{\der\omega_c}=\Lam\omega_c+eD_c,\label{cIc}\\
D_c&=&\int d^3x Z_a^cK_{ab}J_b,\\
\Lam\delta_{cd}&=&\int d^3x Z_a^cK_{ab}Z_b^d,
\eeqa
where $\Lam$ is the moment of inertia of the Skyrmion. The Lagrangian 
is simply 
transformed into the Hamiltonian as 
\beqa
H_{\rm Sky}&=&I_a\omega_a-L=\frac{I_a(I_a-eD_a)}{\Lam}
-L_{\rm Sky}\noeq\\
&=&M_S+\frac{I_a^2}{2\Lam}-\frac{eI_aD_a}{\Lam}
+e^2\left[\frac{C_a^2}{\Lam}-
\int d^3xJ_aK_{ab}J_c\right]-\half\int d^3xeA_0B_0.
\label{SkyH}
\eeqa
We observe that the seagull terms   given as 
\beq
H_{\gamma\gamma}^{\rm Sky}
=\frac{e^2}{2\Lam}\left(\int d^3x J_bK_{bc}Z_b^a\right)^2
-\frac{e^2}{2}\int d^3x J_aK_{ab}J_b \label{Sky}
\eeq
could survive depending on the form of $A_0$ in $J_a$.  
This result is used in  previous calculations of the electric 
polarizability in the Skyrme model\Ref{\ref{1}-\ref{6},\ref{SC}}. 
\par
\section{Fluctuations and the vanishing seagull terms}
Now, we introduce the fluctuation fields around the Skyrmion 
configuration as
\beq
\pha=\psi_a(\vx,t)+\chi_a(x),
\eeq
where $\chi_a$ is defined in the laboratory system. Thus the 
time-derivative is given as 
\beq
\dot\pha(x)=Z_a^c\omega_c+\dot \chi_a.
\eeq
We note that the time-derivative of the Skyrmion field is of \Nc{-1/2}, 
while that of the fluctuation is of \Nc{0}.
 Then, the Lagrangian is rewritten as 
\beq
L=\half\int d^3x\left\{(Z_a^c\omega_c+\dot\chi_a+eJ_a)K_{ab}
(Z_b^d\omega_d+\dot\chi_b+eJ_b)-{\cal{V}}+\half eA_0B_0\right\},
\eeq
from which we derive the momenta conjugate to $\omega_c$ and 
$\chi_a$ as 
\beqa
I_c&=&\left(\Lam+\int d^3xZ_a^cK_{ab}\dot\chi_b\right)\omega_c
+eD_c,\label{Ic}\\
\Pi_a&=&K_{ab}(Z_b^d\omega_d+\dot\chi_b+eJ_b),\label{Pia}
\eeqa
respectively. 
There is a linear relation between $I_c$ and $\Pi_a$, the primary 
constraint, 
\beq
F_c=I_c-\int d^3xZ_a^c\Pi_a=0.\label{Fc}
\eeq
The Hamiltonian is formally written as follows:
\beqa
H&=&M_S+\half\int d^3x\Pi_aK^{-1}_{ab}\Pi_b-\int d^3x\Pi_aJ_a
-\half\int d^3xeA_0B_0
\noeq\\
&&+\mbox{ terms of $\chi$'s and their spatial derivatives.}
\eeqa
It should be noticed that this Hamiltonian has no seagull terms. This is 
due to the constraint Eq.(\ref{Fc}), where $I_c$ is expressed in terms of 
$\Pi$ and $Z$ in contrast to Eqs.(\ref{cIc}) and (\ref{Ic}), which contain 
$eD_c$.
\par
Since there are  the constraints among $I_c$'s and $\Pi_a$'s,  we  
impose   the standard  gauge-fixing condition that $\chi_a$'s are  
orthogonal to  the zero-mode wave functions, $Z_a^c$, 
\beq
G_c=\int d^3xZ_a^cK_{ab}\chi_b=0.
\eeq
And, according to the standard method, we decompose  $\Pi_a$ into  the 
transverse component, $\pi_a$,  and  the parallel one to  the zero-mode 
wave functions, where the primary constraint is rewritten as 
\beq
F'_c=\int d^3xZ_a^c\pi_a=0, 
\eeq 
and  then we obtain 
\beq
\Pi_a=\pi_a+\frac{1}{\Lam}K_{ab}Z_b^dI_d.
\eeq 
Following  this  transformation $I_a$ should also be transformed, but 
we can take the same $I_a$ as the conjugate momentum to the rotational
angle {\it in the tree approximation}, because the difference is due 
to the fluctuation fields\Ref{\ref{HSUold}}. 
Finally we find the Hamiltonian satisfying  all the constraints as 
\beqa
H&=&M_s+\frac{I_a^2}{2\Lam}-\frac{eI_aD_a}{\Lam}
-e\int d^3x(\pi_aJ_a+\half A_0B_0)\noeq\\
&&+\half\int d^3x\pia K^{-1}_{ab}\pib+\mbox{power terms of $\chi$}.
\eeqa
The seagull terms do not reappear.  The same conclusion is also 
obtained for the fluctuation introduced in the intrinsic frame of the 
Skyrmion. 
\par
This Hamiltonian is, however, different from the one obtained in 
\Ref{\ref{MW}}, where $\pi_a=K_{ab}\dot\chi_b$ is assumed and 
$\pi_{\rm coll}$ is identified to the remainder of Eq.({\ref{Pia}), that is, 
$\Pi_a$ is decomposed into three conponents, perpendicular, parallel and 
irrelevant to the zero-mode wave function. If one does so, one can easily 
see that $\pi_a$ cannot be the canonical conjugate  momentum of
$\chi_a$: Since
\beq
\int d^3x\Pi_a\dot\chi_a=\int d^3x(\pi_a+eK_{ab}J_b)\dot\chi_a
\eeq
in the Legendre transformation, the momentum field canonically 
conjugate to $\chi_a$ is to be $\tilde\pi_a=\pi_a+K_{ab}J_b$, where 
used is $\int Z_a^cK_{ab}\dot\chi_b=\int Z_a^c\pi_a=0$, 
that is valid only for the tree approximation. Thus, in order to get the 
canonically  conjugate momentum to $\chi_a$,  one has to decompose 
$\Pi_a$ into the two components,  perpendicular to and parallel to 
$Z_a^c$  without any third component in the standard gauge-fixing 
conditions. Since there is no third component like  $eJ_a$, $\Pi_a^2$ 
cannot have the relevant term $e^2J_aK_{ab}J_b$. 
\par
Thus, we are led to the conclusion that the seagull terms do not really 
survive in the Hamiltonian, if we introduce the fluctuations around the 
classical soliton configuration.
\par
\section{Conclusions and discussion}
If we truncate the dynamical variables at  the collective ones in the 
Lagrangian,  the seagull terms could  survive in the Hamiltonian. 
It is known that the surviving  seagull terms give the electric 
polarizability of the nucleon $\alpha_N$ and magnetic one $\beta_N$  
a relation $\alpha_N\cong-2\beta_N$ in contrast to the relation 
$\alpha_N=10\beta_N$ in the chiral perturbation theory at the chiral 
limit\Ref{\ref{chpt}}. $\alpha_N$ from the seagull terms is 
larger than the one in the latter by a factor 3\Ref{\ref{5}}. 
The factor 3 is interpreted as the hidden contribution from the 
degenerate delta isobar\Ref{\ref{5}}, but it  would be difficult to 
apply the same interpretation to the magnetic polarizability, since the 
delta contribution is explicitly added to the magnetic seagull term to 
reduce the net magnetic polarizability. We wonder whether the factor 3  
is a strong evidence of the validity of the surviving  seagull terms. 
Moreover, the seagull terms are not stable against any flucutations 
around the Skyrmion as shown above. 
\par
If we want to go  beyond the collective variables and to calculate loop
corrections to the polarizabilities and other quantities in the Skyrme 
model, we have to introduce the fluctuation fields in the Lagrangian, and 
then we have to start with the vanishing  seagull terms  
for the electric polarizability.  In this respect we  point out that 
there are also no direct two-photon couplings to the nucleon 
contributing to the electric polarizability of \Nc{} in the chiral 
perturbation theory with the nucleon field\Ref{\ref{chpt}}. 
Since we cannot attribute the electric polarizability to the seagull 
terms, we have to search for other  origin of the electric polarizabilty 
of \Nc{}. We have performed this task by taking account of the 
contributions to the Compton scattering amplitude from  the dispersion 
integrals of the low energy  pion-photoproduction ones obtained 
within the Skyrme model\Ref{\ref{SUprd}}.  We are able to take into 
account the nucleon and delta mass difference of \Nc{-1}; if we discard 
the contributions from the delta isobar, we have the same 
electromagenetic polarizabilities  as  those by the 
one-loop calculations in the chiral perturbation theory at the chiral 
limit. Thus,  it is possible in the Skyrme model to give reasonable 
values to the electromagnetic polarizabilities of \Nc{} by pion-loop 
calculations even if the classical contributions do not exist. 

\newpage 
\baselineskip 24pt
\begin{center}
{\bf References}
\end{center}
\def\labelenumi{[\theenumi]}
\def\Ref#1{[{#1}]}
\def\npb#1#2#3{{Nucl. Phys.\,}{\bf B{#1}}\,(#3), #2}
\def\npa#1#2#3{{Nucl. Phys.\,}{\bf A{#1}}\,(#3),#2}
\def\plb#1#2#3{{Phys. Lett.\,}{\bf B{#1}}\,(#3),#2}
\def\prd#1#2#3{{Phys. Rev.\,}{\bf D{#1}}\,(#3),#2}
\begin{enumerate}
\item \label{1} E. M. Nyman, \plb{142}{388}{1984}.
\item \label{2} M. Chemtob, \npa{473}{613}{1987}.
\item \label{3} N. N. Scoccola and W. Weise, \plb{232}{278}{1989}:  \\
\npa{517}{495}{1990}.
\item \label{4} S. Scherer and P. J. Mulders, \npa{549}{521}{1992}.
\item \label{5} W. Broniowsky, M. K. Banerjee and T. D. Cohen, 
\plb{283}{22}{1992}.\\
W. Broniowsky and T. D. Cohen, \prd{47}{299}{1993}.\\
B. Golli and R. Sraka, \plb{312}{24}{1993}.
\item \label{6} C. Gobbi, C. L. Schat and N. N. Scoccola, 
\npa{598}{318}{1996}.
\item \label{Lvov} A. I. L'vov, Int. J. Mod. Phys. {\bf A30}(1993),5267.
\item \label{SUplb} S. Saito and M. Uehara, \plb{325}{20}{1994}.
\item \label{SUprd} S. Saito and M. Uehara, \prd{51}{6059}{1995}.
\item \label{SC}  N. N. Scoccola and T. D. Cohen, \npb{596}{599}{1996}.
\item \label{MW} F. Meier and H. Walliser, hep-ph/9602359.
\item \label{HSUnew} A. Hayashi, S. Saito and M. Uehara, 
\prd{46}{4856}{1992}.
\item \label{HSUold} A. Hayashi, S. Saito and M. Uehara, 
\prd{43}{1520}{1991}.
\item \label{chpt} V. Bernard, N. Kaiser and U. -G. Meissner, 
\npb{373}{346}{1992}.\\
V. Bernard, N. Kaiser, J. Kambor and U.-G. Meissner, 
\npb{388}{315}{1992}.
\end{enumerate}
\end{document}